\documentclass[useAMS,usenatbib]{mn2e}

\usepackage{natbib}
\usepackage{graphicx}
\usepackage{multicol}
\usepackage{longtable}
\usepackage{amssymb}
\usepackage[width=1.0\textwidth]{caption}
\usepackage[usenames,dvipsnames]{color}
\RequirePackage[pdftex, plainpages = false, pdfpagelabels]{hyperref}
\definecolor{Red}{rgb}{1.0,0,0}

\newcommand{\etal}    {{\it et al}}                          
\newcommand{\AS}    {{\sc Autostructure}}
\newcommand{\Ss}    {{\sc Superstructure}}
\newcommand{\Ii}      {~{\sc i}}
\newcommand{\II}      {~{\sc ii}}
\newcommand{\III}     {~{\sc iii}}

\newcommand{\ps} [1]{\overline{#1}}

\title[Atomic data for {\rm Co\II} infrared lines]
{Collision strengths and transition probabilities for Co\II\ infrared forbidden lines}
\author[P.J. Storey \& C.J. Zeippen \& T. Sochi]
{P.J. Storey$^{1}$\thanks{E-mail: pjs@star.ucl.ac.uk}, C.J. Zeippen$^{2}$, Taha Sochi$^{1}$ \\
$^{1}$Department of Physics and Astronomy, University College London, Gower Street, London WC1E
6BT, UK \\ $^{2}$LERMA, Observatoire de Paris, ENS, UPMC, UCP, CNRS, 5 Place Jules Janssen, F-92195
Meudon Cedex, France}

\makeatletter \DeclareRobustCommand{\element}[1]{\@element#1\@nil}
\def\@element#1#2\@nil{%
  #1%
  \if\relax#2\relax\else\MakeLowercase{#2}\fi}
 \makeatother

\begin{document}

\date{Accepted 2015 November 24.  Received 2015 November 24; in original form 2015 September 10}

\maketitle

\label{firstpage}

\begin{abstract}

\noindent We calculate collision strengths and their thermally-averaged Maxwellian values for
electron excitation and de-excitation between the fifteen lowest levels of singly-ionised cobalt,
Co$^+$, which give rise to emission lines in the near- and mid-infrared. Transition probabilities
are also calculated and relative line intensities predicted for conditions typical of supernova
ejecta. The diagnostic potential of the 10.52, 15.46 and 14.74~$\mu$m transition lines is briefly
discussed.
\end{abstract}


\section{Introduction} \label{Introduction}

Cobalt transition lines are useful astronomical probes especially in analysing the observational
data of supernovae (SNe) where these emissions can be used to assess the nucleosynthesis and decay
processes associated with the SN explosion. The spectra of supernova ejecta show prominent infrared
forbidden lines of singly- and doubly-ionised ions of the iron group elements: nickel, cobalt and
iron, where the formation of these elements is largely based on the radioactive decay chain
$^{56}$Ni $\rightarrow ^{56}$Co $\rightarrow ^{56}$Fe \citep{ColgateM1969, KuchnerKPL1994,
BowersMGWPe1997, LiuJSQSP1997, ChurazovSIKJe2014}. Hence, the lines of these elements can be used
to examine the underlying nuclear processes. These lines can also be used for the determination of
particle number density and for analysing the thermodynamic conditions of these objects
\citep{NussbaumerS1988, KuchnerKPL1994, BowersMGWPe1997, AdelmanGL2000, PelosoCSM2005,
BergemannPG2010}.

Observations of cobalt lines in super and symbiotic novae, largely forbidden transitions in the
infrared and optical parts of the spectrum, have been reported in a number of studies
\citep{AxelrodThesis1980, LiMS1993, JenningsBWM1993, KuchnerKPL1994, Dinerstein1995,
BowersMGWPe1997, LiuJSQSP1997, McKennaKHPRe1997, ChurazovSIKJe2014}. Definite or tentative
observations of certain forbidden transitions of Co in the spectra of planetary nebulae and H\II\
regions have also been reported \citep{BaluteauZMP1995, SharpeeWBH2003, EstebanPRRPR2004,
ZhangLLPB2005, PottaschS2005b, SharpeeZWP2007, WangL2007, FangL2011}. Cobalt lines have also been
observed in other astronomical objects like early-type stars \citep{AdelmanGL2000} and cool stars
\citep{BergemannPG2010} as well as in the solar spectrum \citep{KerolaA1976, SalihLW1985,
PickeringRUJ1998, BergemannPG2010}. Relevant observational data related to the SN spectra in
general and to the IR lines in particular have been gathered in the past using various instruments
such as the Infrared Spectrograph (IRS) on the Spitzer Space Telescope and Himalaya Faint Object
Spectrograph Camera (HFOSC) on the Himalayan Chandra Telescope (see e.g. \citet{SahuASM2006} and
\citet{JerkstrandFMSES2012}).

There are several experimental investigations (see e.g. \citet{SugarC1981, SugarC1985, SalihLW1985,
UrrutiaUNJ1994, MullmanCL1998, MullmanLZF1998,PickeringRUJ1998, RuffoniP2013}) in which atomic
data, such as energy levels and radiative transition probabilities, have been reported.

\citet{NussbaumerS1988} reported calculations of forbidden transition probabilities between the
energetically lowest eight levels of Co$^+$ using a multi-configuration atomic model. The
calculations, which are based on employing the \Ss\ code \citep{EissnerJN1974, NussbaumerS1978},
have also included summarisation of similar calculations for Co\Ii\ and Co\III\ transitions. These
data were intended to facilitate the analysis of the observations of SNe in general and SN1987A in
particular.
\citet{RaassenPU1998} conducted experimentally-based calculations of oscillator strengths and
transition probabilities for a number of dipole-allowed and dipole-forbidden transitions of
singly-ionised cobalt using a semi-empirical approach based on employing an orthogonal operator method
within an intermediate coupling scheme. The data were intended for use in analysing astronomical
spectra from such objects as Co stars and late type supernovae.
For similar purposes, \citet{Quinet1998} computed radiative transition probabilities for the
dipole-forbidden transitions between the 47 metastable energy levels in the 3d$^8$, 3d$^7$4s and
3d$^6$4s$^2$ configurations of the singly-ionised cobalt using a configuration interaction
Hartree-Fock relativistic approach with optimised radial parameters based on the available observed
energy levels.

In this paper, we report computational atomic data in the form of collision strengths and their
thermally-averaged Maxwellian values for electron excitation and de-excitation between the 15
lowest levels of Co$^+$ as well as transition probabilities for a number of the forbidden lines of
Co\II\ in the infrared section of the spectrum. The investigation is generally based on employing
the R-matrix method and codes\footnote{{See Badnell: R-matrix write-up on WWW. URL:
amdpp.phys.strath.ac.uk/UK\_RmaX/codes/serial/WRITEUP.}} \citep{BerringtonBCCRT1974,
BerringtonBBSSTY1987, HummerBEPST1993, BerringtonEN1995} where the scattering calculations are
conducted using a 13-configuration atomic target within a Breit-Pauli intermediate coupling
approximation.

In Section \ref{AtomicStructure} of the present paper, we discuss the target used in our Co$^+$
model and give the resulting transition probabilities. Details of the Breit-Pauli R-matrix electron
scattering calculations can be found in Section \ref{Scattering}. Results and some diagnostics
appear in Section \ref{Results}. The paper is finalised in Section \ref{Conclusions} with general
conclusions and discussion.

\section{\element{Co}$^+$ Atomic Structure} \label{AtomicStructure}

\subsection{The scattering target}

\begin{figure}
\centering
\includegraphics[height=7cm, width=8.5cm]{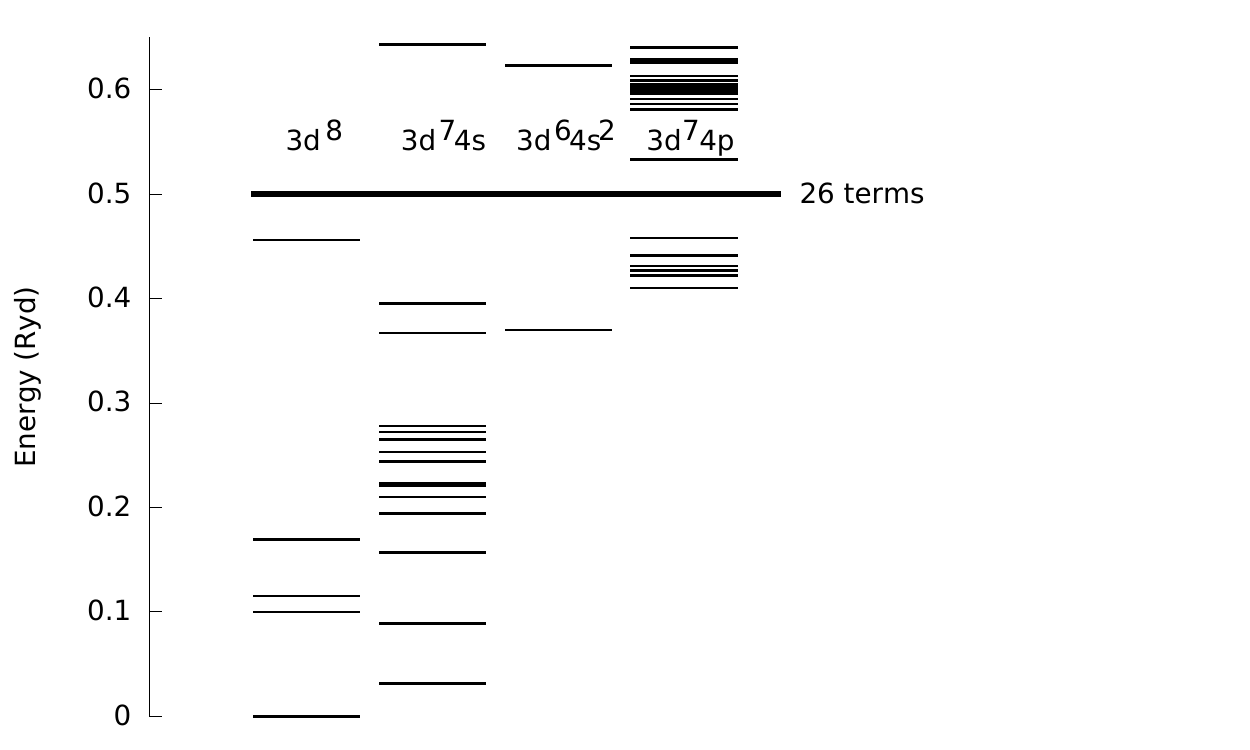}
\caption[]{Schematic term energy diagram of Co~{\sc ii}. The solid line shows the extent of the
close-coupled target states.} \label{termdiagram}
\end{figure}

\begin{figure}
\centering
\includegraphics[height=8cm, width=6.5cm]{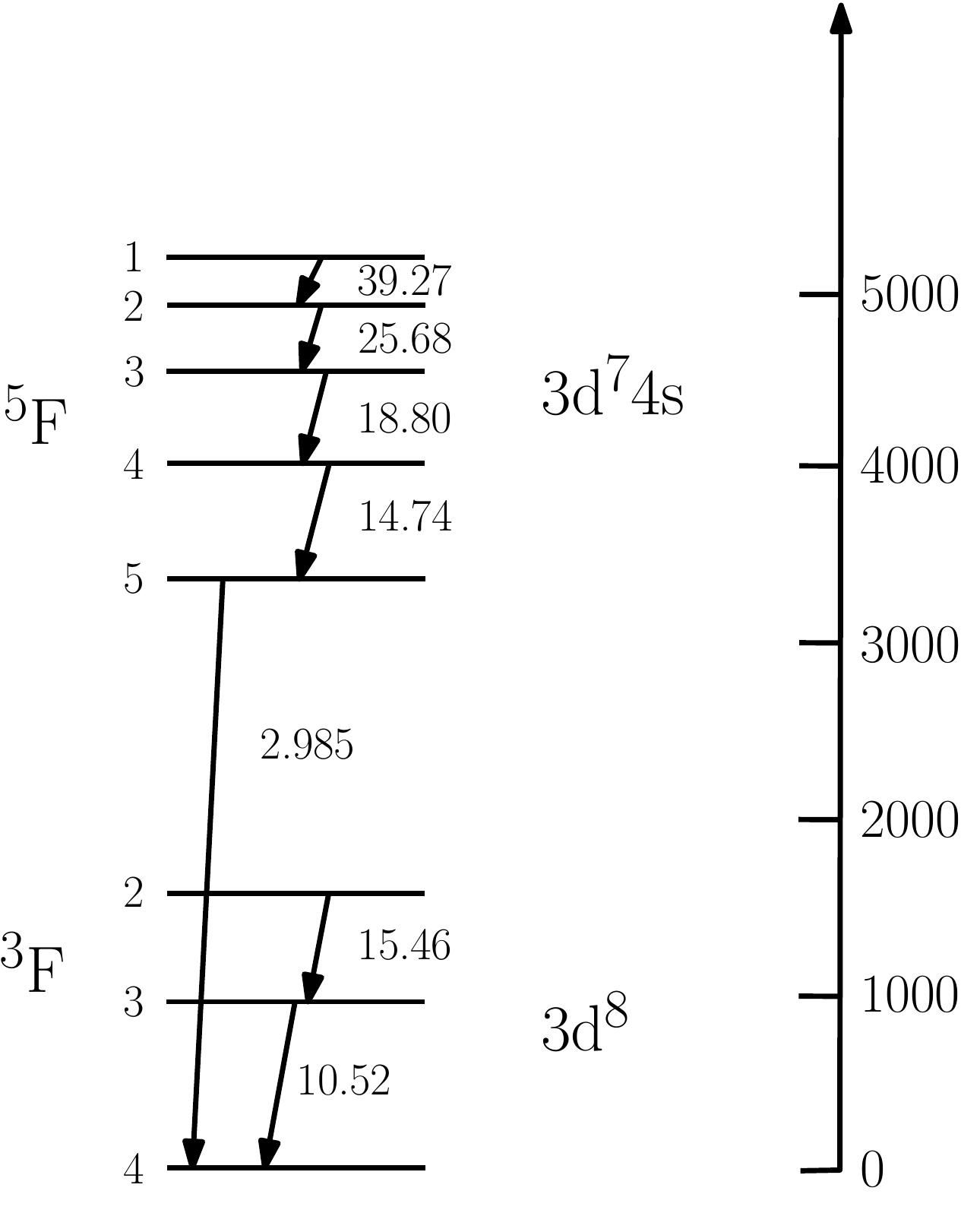}
\caption{The lowest levels of Co$^+$ and the mid-infrared transitions between them. The energy
scale on the right is in cm$^{-1}$ while the transition wavelengths next to the arrows are in
$\mu$m.} \label{levelsandirlines}
\end{figure}

A schematic diagram of the term structure of Co~{\sc ii} is shown in Figure~\ref{termdiagram}. The
lowest 19 terms are of even parity from the configurations 3d$^8$ and 3d$^7$4s. In this work we are
concerned with the excitation mechanisms of the forbidden transitions among these even-parity
terms, particularly those lying in the near- and mid-infrared. The levels giving rise to
mid-infrared transitions and the corresponding wavelengths are shown in
Figure~\ref{levelsandirlines}. Transitions from higher terms give rise to near-infrared lines but
these are expected to be weaker at the temperatures typical of supernova ejecta.

It is expected that the odd-parity terms of the 3d$^7$4p configuration will give rise to resonance
series that affect the collision strengths for excitation of the low-lying even-parity levels and
should therefore be included in the target for the electron scattering calculation. The extent of
our target is shown by the solid line in Figure~\ref{termdiagram} and includes 26 terms and 72
levels. The target states were expanded over the set of thirteen electron configurations listed in
Table~\ref{configs} and the target wavefunctions were calculated with the program \AS,
\citep{EissnerJN1974, NussbaumerS1978, AS2011}, which uses radial wavefunctions calculated in
scaled Thomas-Fermi-Dirac statistical model potentials.  The scaling parameters were determined by
minimising the sum of the energies of all the target terms, computed in $LS$-coupling, {\it i.e.}
neglecting all relativistic effects. The resulting scaling parameters are given in
Table~\ref{scale} where a negative scaling parameter, $\lambda_{nl}$, signifies a hydrogenic
correlation orbital with nuclear charge 27$|\lambda_{nl}|$.

\begin{table}
\caption{The target configuration basis where the core structure is suppressed. The bar indicates a
correlation orbital.}
\begin{flushleft}
\centering
\begin{tabular}{llll}
\noalign{\hrule}
 & 3d$^8$          & 3d$^7$\,4p  \cr
 & 3d$^7$\,4s      & 3d$^6$\,4s\,4p \cr
 & 3d$^6$\,4s$^2$  &  \cr
 & 3d$^6$\,4p$^2$  & \cr
 &                 & \cr
 & 3d$^7$\,$\ps4$d & \cr
 & 3d$^7$\,$\ps5$s      & \cr
 & 3d$^7$\,$\ps5$d      & \cr
 & 3d$^7$\,$\ps6$d      & \cr
 & 3d$^6$\,4s\,$\ps4$d  &   \cr
 & 3d$^6$\,$\ps4$d$^2$  &   \cr
 & 3d$^5$\,4s$^2$\,$\ps4$d  & \cr
\noalign{\hrule}
\end{tabular}
\end{flushleft}
\label{configs}
\end{table}

\begin{table}
\caption{Potential scaling parameters. The bar over the principal quantum number and the minus sign
attached to the value of a scaling parameter signifies a correlation orbital.}
\centering %
\begin{tabular}{lr@{\hskip 1.3cm}lr@{\hskip 1.3cm}lr}
\noalign{\hrule}
 1s       & 1.43187 & 2p & 1.08106 & 3d      & 1.02832  \cr
 2s       & 1.13734 & 3p & 1.04220 & $\ps4$d & -0.34647 \cr
 3s       & 1.05922 & 4p & 0.98776 & $\ps5$d & -2.19279 \cr
 4s       & 0.98959 &    &         &                $\ps6$d & -2.00375 \cr
 $\ps5$s  & -0.55000   &    &         &         &          \cr
\noalign{\hrule}
\end{tabular}
\label{scale}
\end{table}

In Table~\ref{termlist} we compare the term energies calculated with our target with experiment for
the 26 terms of the target. The calculated term energies include one-body relativistic shifts, the
mass and Darwin terms, and the spin-orbit interaction. This is the level of approximation that
applies in the R-matrix code used for the electron scattering calculation. In Table~\ref{levellist}
we compare the calculated energies of the 15 lowest levels with experimental values. In this table
we also show the results obtained when the two-body fine structure interactions described by
\citet{EissnerJN1974} are included. The overall fine-structure splittings of the tabulated terms
are significantly improved when the two-body terms are included, with the average difference from
experiment falling from 5.3\% to 1.8\%. Table~\ref{levellist} serves as a key to the levels for use
in later results for transition probabilities, collision strengths and effective collision
strengths.

\begin{table}
\caption{Energies of the 26 target terms in cm$^{-1}$. The calculated values include only the
spin-orbit contribution to the fine-structure energies. Core structure is suppressed from all configurations.}
\centering %
\begin{tabular}{lllll}
\noalign{\hrule}
 & & \hspace{1.5cm} & \multicolumn{2}{c}{\hspace{-2cm}Term Energy} \cr
 & Config. & Term & Exp.$^{\dagger}$ & Calc.  \cr
\noalign{\hrule}
 & 3d$^8$   & $^3$F &     0   &   0   \\
 & 3d$^7$4s & $^5$F &  3457   & 2961  \\
 &          & $^3$F &  9774   & 9570  \\
 & 3d$^8$   & $^1$D & 10954   & 13639  \\
 &          & $^3$P & 12648   & 15349  \\
 & 3d$^7$4s & $^5$P & 17275   & 20043  \\
 & 3d$^8$   & $^1$G & 18277   & 20935  \\
 & 3d$^7$4s & $^3$G & 21261   & 23588  \\
 &          & $^3$P & 23479   & 26652  \\
 &          & $^3$P & 24441   & 29699  \\
 &          & $^1$G & 24450   & 27245  \\
 &          & $^3$H & 26747   & 29406  \\
 &          & $^1$P & 26888   & 33570  \\
 &          & $^3$D & 27353   & 31649  \\
 &          & $^1$H & 29870   & 32672  \\
 &          & $^1$D & 30502   & 35335  \\
 & 3d$^7$4s & $^3$F & 40228   & 46895  \\
 & 3d$^6$4s$^2$ & $^5$D  &    40560$^a$ & 41967  \\
 & 3d$^8$   & $^1$S &  42303  & 51388 \\
 & 3d$^7$4s & $^1$F & 43394   & 50213  \\
 & 3d$^7$4p & $^5$F$^{\rm o}$  & 45018 & 42614 \\
 &          & $^5$D$^{\rm o}$  & 46319 & 44049 \\
 &          & $^5$G$^{\rm o}$  & 46905 & 44577 \\
 &          & $^3$G$^{\rm o}$  & 48507 & 46823  \\
 &          & $^3$F$^{\rm o}$  & 49518 & 47843 \\
 &          & $^3$D$^{\rm o}$  & 51289 & 49868 \\
\noalign{\hrule} \multicolumn{5}{l}{$^{\dagger}$Experimental energies are from \citet{SugarC1985}.} \\
\multicolumn{5}{l}{$^a$The $^5$D$_0$ energy, which is not known experimentally, was} \\
\multicolumn{5}{l}{derived from the calculated $^5$D$_1$--$^5$D$_0$ energy separation.}
\end{tabular}
\label{termlist}
\end{table}

\begin{table}
\caption{Energies in cm$^{-1}$ of the 15 lowest target levels.} \centering
\begin{tabular}{lrlcrrr}
\noalign{\hrule}
 & Index & Config. & Level &
\multicolumn{1}{c}{Exp.$^{1}$} & Calc.$^2$ &  Calc.$^3$ \\
\noalign{\hrule}
       &  1  &  3d$^8$            & $a^3$F$_4$      & 0. & 0. & 0. \\
       &  2  &                    & $a^3$F$_3$      & 951  & 988 & 972 \\
       &  3  &                    & $a^3$F$_2$     &  1597  & 1676 & 1645 \\
       &  4  &  3d$^7$4s          & $a^5$F$_5$     & 3351  & 2899 & 2876 \\
       &  5  &                    & $a^5$F$_4$     & 4029  & 3562 & 3550 \\
       &  6  &                    & $a^5$F$_3$     & 4561  & 4088 & 4082 \\
       &  7  &                    & $a^5$F$_2$     & 4950  & 4476 & 4474 \\
       &  8  &                    & $a^5$F$_1$     & 5205  & 4732 & 4731 \\
       &  9  & 3d$^7$4s           & $b^3$F$_4$     & 9813  & 9642 & 9623 \\
       & 10  &                    & $b^3$F$_3$     & 10709 & 10531 & 10523 \\
       & 11  &                    & $b^3$F$_2$     & 11322 & 11156 & 11153 \\
       & 12  & 3d$^8$             & $a^1$D$_2$     & 11651 & 14367 & 14378 \\
       & 13  &                    & $a^3$P$_2$     & 13261 & 16022 & 15970 \\
       & 14  &                    & $a^3$P$_1$     & 13404 & 16094 & 16094 \\
       & 15  &                    & $a^3$P$_0$     & 13593 & 16305 & 16297 \\
\noalign{\hrule}
\multicolumn{6}{l}{$^{1}$ \citealt{SugarC1985}.} \\
\multicolumn{6}{l}{$^2$ Calculated with only spin-orbit interaction.} \\
\multicolumn{6}{l}{$^3$ As 2 plus two-body fine-structure interactions.} \\
\end{tabular}
\label{levellist}
\end{table}

A further measure of the quality of the target is a comparison between weighted oscillator
strengths, $gf$, calculated in the length and velocity formulations. Good agreement between the two
formulations is a necessary but not sufficient condition for ensuring the quality of the target
wavefunctions. This comparison, given in Table~\ref{gflv}, shows an average difference in the
absolute values of $gf$ of 10.9\% between the two formulations, which we consider acceptable for an
open d-shell system.

\begin{table}
\caption{Weighted $LS$ oscillator strengths, $gf$, in the length and velocity formulations from the
energetically lowest three terms.} \centering
\begin{tabular}{llllrr@{\hskip 1.5cm}rr}
\noalign{\hrule}
 & \multicolumn{5}{c}{\hspace{-1.5cm}Transition}  & $gf_{_L}$ & $gf_{_V}$ \\
\noalign{\hrule}
 & 3d$^8$    & $^3$F & -- & 3d$^7$4p & $^3$G$^{\rm o}$ & 0.17  &  0.21      \\
 &           &       & -- &         & $^3$F$^{\rm o}$ & 1.34 & 1.55 \\
 &           &       & -- &         & $^3$D$^{\rm o}$ & 0.75 & 0.81  \\
 & 3d$^7$4s  & $^5$F & -- & 3d$^7$4p & $^5$F$^{\rm o}$ & 10.75 & 10.70 \\
 &           &       & -- &         & $^5$D$^{\rm o}$ & 7.30 & 6.62 \\
 &           &       & -- &         & $^5$G$^{\rm o}$ & 14.23 & 14.97 \\
 & 3d$^7$4s  & $^3$F & -- & 3d$^7$4p & $^3$G$^{\rm o}$ & 7.51  &  9.09      \\
 &           &       & -- &         & $^3$F$^{\rm o}$ & 5.82  & 6.52 \\
 &           &       & -- &         & $^3$D$^{\rm o}$ & 4.29  & 4.42  \\
\noalign{\hrule}
\end{tabular}
\label{gflv}
\end{table}

\subsection{Transition probabilities}\label{Avalues}

Using the target wavefunctions described above, we computed the forbidden transition probabilities
among the low-lying even parity terms. The calculated energies are replaced by experimental
energies to correct the energy factors connecting the {\it ab initio} calculated line strengths to
the transition probabilities. The results are given in Table~\ref{Avalues2} with comparison to
values obtained from previous investigations. We only tabulate those probabilities from a given
upper level which exceed 1\% of the total probability from that level. We also tabulate the
magnetic dipole transition probabilities between the levels of the $^3$F and $^5$F terms calculated
assuming the states are described by pure LS-coupling, in which case the line strength is given by
a simple formula \citep{NussbaumerS1988}.

The mid-infrared transitions between the levels of the 3d$^8$~$a^3$F and 3d$^7$4s~$a^5$F terms are
dominated by magnetic dipole decays which change $J$ by unity, with electric quadrupole transitions
being orders of magnitude smaller. This leads to a step-wise decay through the levels within each
term, as illustrated in Figure~\ref{levelsandirlines}. The lowest $a^5$F level, $a^5$F$_5$, decays
by a very weak magnetic dipole transition to the ground $a^3$F$_4$ level. We discuss this
transition in more detail at the end of this section.

We find very close agreement with \citet{NussbaumerS1988} for transition probabilities within the
$a^3$F and $a^5$F terms, with differences less than 1\% between the two calculations. Differences
of up to 45\% occur for transitions between the $b^3$F and two lower terms and within that term.

We also find reasonable agreement with the work of \citet{Quinet1998} for the transitions of
interest here, the magnetic dipole transition probabilities within the $a^3$F and $a^5$F terms. His
results differ by no more than 20\% from the present work for those transitions. Larger
differences, of up to 60\% are seen for transitions from the higher $b^3$F, $a^1$D and $a^3$P terms
to the ground $a^3$F term, although the remaining transitions show much smaller differences,
typically about 10\%. The calculation of \citet{Quinet1998} uses a similar configuration basis to
the present work but employs a Hartree-Fock approach which incorporates fitting of Slater
parameters to experimental energies. Consequently, that method yields calculated energies in much
better agreement with experiment than our {\it ab initio} results. We do, however, make empirical
corrections to the $LS$-Hamiltonian matrix to bring our final calculated level energies into good
agreement with experiment, to ensure that the spin-orbit interactions between levels of different
terms are corrected for any errors due to incorrect term energy separations. We also use
experimental energies in the calculation of transition probabilities from line strengths.

Table~\ref{Avalues2} also shows the forbidden transition probabilities calculated by
\citet{RaassenPU1998}, which show very good agreement with the present work, differing by less than
10\% in all but three cases and agreeing within 1\% for all the magnetic dipole transitions between
the levels of the $a^3$F and $a^5$F terms. Despite this very close agreement for the infrared
transitions, the larger differences between our results and those of \citet{Quinet1998} should
probably be viewed as a measure of the uncertainty in the values of the transition probabilities
for this rather complex ion.

The $a^5$F$_5$--$a^3$F$_4$, which was briefly referred to above, requires a separate discussion.
Since there are no magnetic dipole matrix elements between different terms, this transition must
proceed {\it via} an interaction between 3d$^8~a^5$F$_4$ and 3d$^7$4s~$a^3$F$_4$. As discussed by
\citet{NussbaumerS1988}, there is no direct spin-orbit interaction between $a^3$F$_4$ and
$a^5$F$_4$ levels; only very small two-body fine-structure terms. A combination of configuration
interaction between $a^3$F and $b^3$F, plus spin-orbit interaction between $b^3$F$_4$ and
$a^5$F$_4$, is also excluded because the matrix element for the $a^3$F -- $b^3$F electrostatic
interaction is also zero. In practice, the transition can occur {\it via} configuration interaction
between $a^3$F and $c^3$F, which is non-zero, and spin-orbit interaction between $c^3$F$_4$ and
$a^5$F$_4$, but the $c^3$F is at relatively high energy (40228~cm$^{-1}$) so the interaction is
very weak.

In their discussion of this transition, \citet{NussbaumerS1988} point out that since it is very
weak, it might be significantly altered if the relativistic corrections to the magnetic dipole
operator described by \citet{EZ81} were included. These corrections are included in the current
work and we find that they change the $a^5$F$_5$--$a^3$F$_4$ transition probability by less than
1\%. Among the four lowest terms, the largest changes due to these corrections, ranging from 0 to
3.2\%, are seen in the $b^3$F--$a^5$F multiplet. We also find that the exclusion of two-body
fine-structure terms only changes the  $a^5$F$_5$--$a^3$F$_4$ transition probability by a small
amount (5\%).

For this transition, our calculated probability of 9.93$\times 10^{-6}$~s$^{-1}$ is a factor 2.0
times larger than the value of \citet{NussbaumerS1988} and a factor 3.2 times larger than that
found by \citet{Quinet1998}.  \citet{RaassenPU1998} do not give a value for this transition
probability. In addition to the mechanism described above, we find that the $a^3$F$_4$ and
$a^5$F$_4$ levels can interact {\it via} $^3$F terms belonging to the 3d$^7$4d configuration. We
believe that it is presence of these interactions that causes the large differences between the
different calculations. The present calculation and that of \citet{NussbaumerS1988} utilise a
short-range correlation 4d orbital to approximate the effect of the infinite series of bound and
continuum d orbitals that should be accounted for. \citet{Quinet1998} does not use such orbitals
and so underestimates the contribution of the 3d$^7$nd configurations to the transition
probability.

Within the $a^3$F and $a^5$F terms, radiative decays occur stepwise between adjacent levels with
probabilities of order $10^{-2}$~s$^{-1}$, while the probability of a transition between the $a^5$F
and $a^3$F terms is typically four orders of magnitude smaller. Hence the only transition of
significance between the two terms is $a^5$F$_5$--$a^3$F$_4$. In a physical situation where levels
are excited by electron collisions and decay by radiative decay or collisional de-excitation this
transition alone determines the critical electron density above which the rate of collisional
de-excitation of the $a^5$F levels is larger than radiative decay. We return to this point in
Section~\ref{irratios}.

\section{Scattering Calculations} \label{Scattering}

The Breit-Pauli R-matrix method which is used in this calculation is described fully elsewhere
\citep{HummerBEPST1993, BerringtonEN1995} and references therein. The calculations described here
were made with the parallel versions of the codes \footnote{{See Badnell: R-matrix write-up on WWW.
URL: http://amdpp.phys.strath.ac.uk/tamoc/codes/parallel/PWRITEUP.}}. We use an R-matrix boundary
radius of 14.9~au, to encompass the most extended target orbital (4p). The expansion of each
scattered electron partial wave is over a basis of 12 functions within the R-matrix boundary, and
the partial wave expansion extends to a maximum of $2J=19$.

\begin{figure}
\centering %
{\begin{minipage}[]{0.5\textwidth} \centering \includegraphics[height=3.5cm, width=7.5cm]
{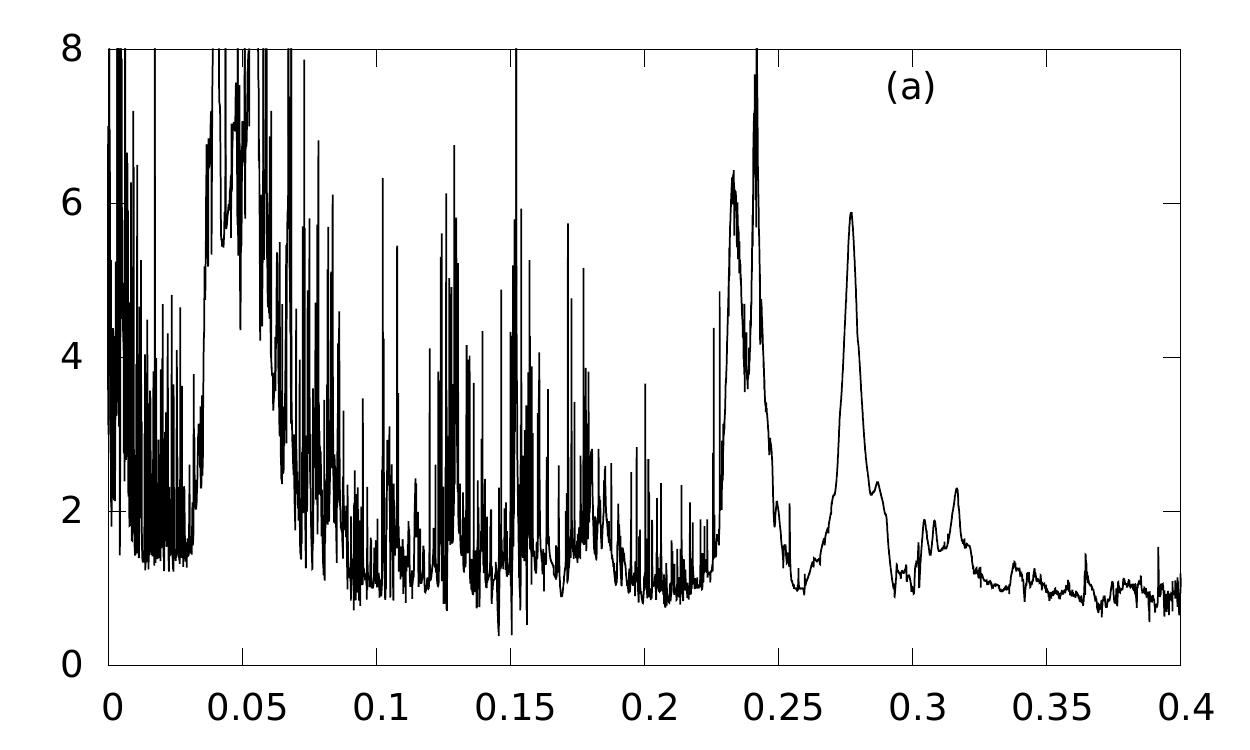}
\end{minipage}} \vspace{0cm}
\centering %
{\begin{minipage}[]{0.5\textwidth} \centering \includegraphics[height=3.5cm, width=7.5cm]
{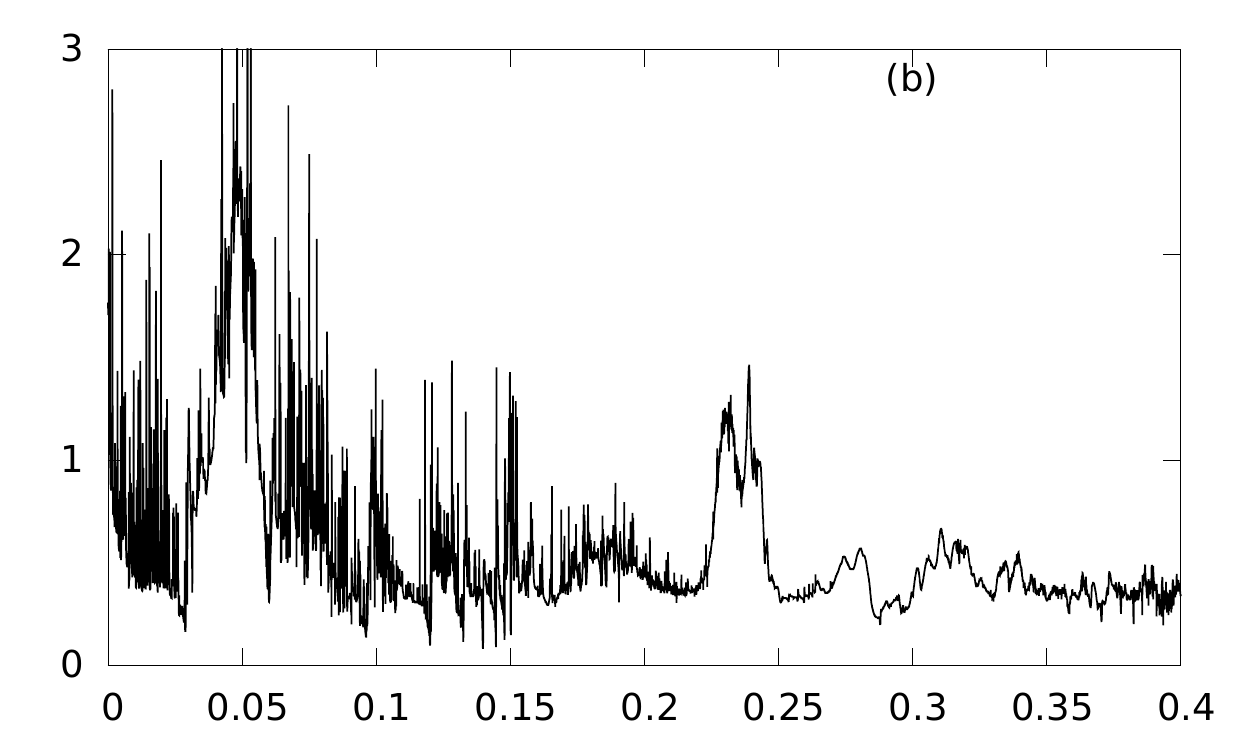}
\end{minipage}} \vspace{0cm}
\centering %
{\begin{minipage}[]{0.5\textwidth} \centering \includegraphics[height=3.5cm, width=7.5cm]
{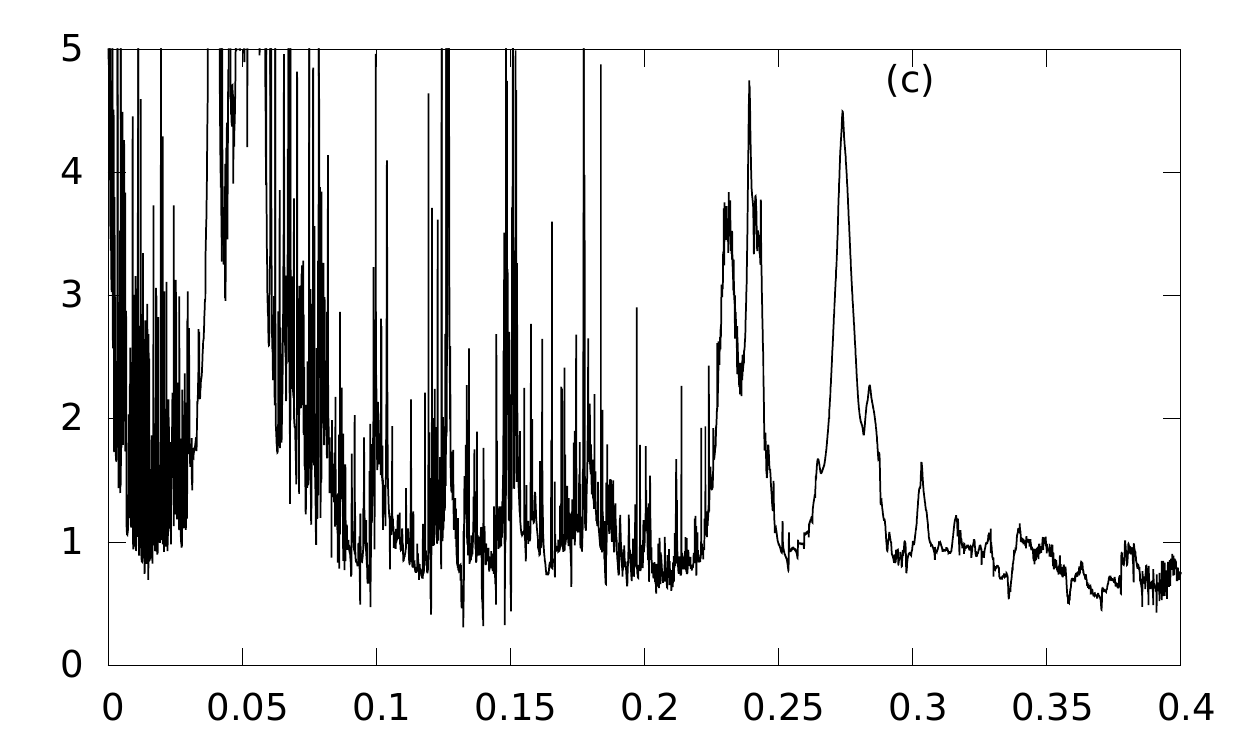}
\end{minipage}}
\caption{Collision strength (vertical axis) versus final electron energy in Rydberg (horizontal
axis) for the (a) 1-2, (b) 1-3 and (c) 2-3 transitions. For level indexing refer to
Table~\ref{levellist}. \label{plotOmega}}
\end{figure}

Collision strengths are computed at 20000 equally spaced values of the energy in the resonance
region and a further 1000 values in the region where all scattering channels are open, up to an
incident electron energy of 1.5~Rydberg. In Figure~\ref{plotOmega} we illustrate our results with
the calculated collision strengths between the three levels of the ground 3d$^8$~$^3$F term as a
function of final electron energy up to 0.4~Rydberg above threshold.  Due to the multiple close
lying thresholds the collision strengths display very dense and complex resonance structures.

\section{Results and Discussion} \label{Results}

In Table~\ref{upstable}, the final thermally averaged collision strengths between the 15
energetically lowest levels are given as a function of electron temperature. In the energy region
where all scattering channels are open we find some small irregular features in the collision
strengths that are almost certainly non-physical and caused by the correlation orbitals in the
target representation. We have computed thermally averaged collision strengths for the transitions
and temperature range given in Table~\ref{upstable} both including and excluding the contribution
from the region of all channels open, and find the largest change for any transition is 0.3\% at
log$_{10}T=4.0$, 2.4\% at log$_{10}T=4.2$ and 9.4\% at log$_{10}T=4.4$. For the excitations from
the $a^3$F levels to the $a^5$F levels, the maximum difference at any temperature is 2.2\%. Hence
for the infrared transitions of interest here the contribution from any non-physical features is
insignificant. The values tabulated in Table~\ref{upstable} were computed using the full energy
range.

\subsection{The IR line ratios}\label{irratios}

\begin{figure}
\centering
\includegraphics[height=6cm, width=8cm]{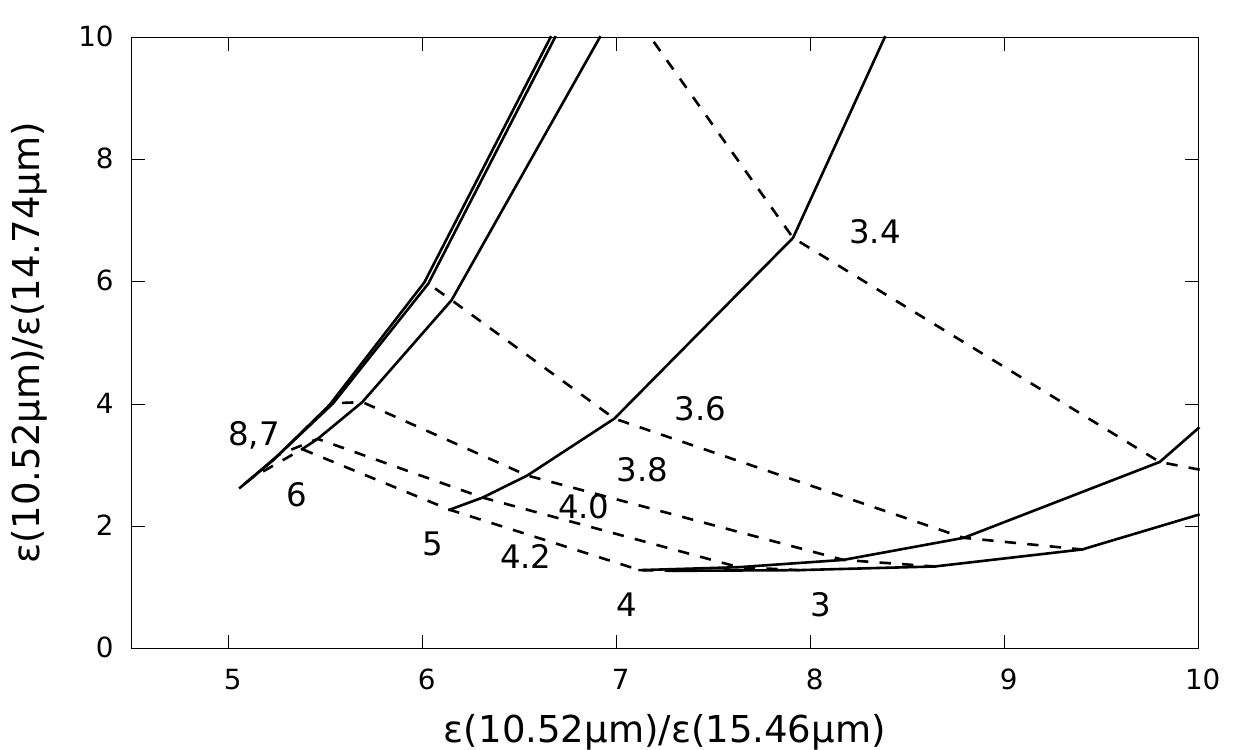}
\caption{Emissivity ratios of pairs of Co~{\sc ii} infrared lines, showing contours of constant
electron temperature (dashed lines) and electron number density (solid lines). Lines are labelled
with the log$_{10}$ of temperature in K or density in cm$^{-3}$.} \label{tworat}
\end{figure}

The transitions between the levels of the $a^3$F and $a^5$F terms give rise to mid-infrared lines
lying between $10.5$ and 39.3~$\mu$m. The relative intensity of the two lines arising from the
$a^3$F term (10.52~$\mu$m and 15.46~$\mu$m, see Figure \ref{levelsandirlines}) is sensitive to
electron density for densities less than about $10^6$~cm$^{-3}$ and relatively insensitive to
temperature for temperatures greater than about $3000$~K due to the low excitation energies. The
lines from within the higher lying $a^5$F term, the 14.74~$\mu$m line for example, are more
sensitive to electron temperature. Ratios of the intensities of these three lines can therefore be
used to simultaneously determine the electron temperature and density of the emitting material if
they lie in the appropriate range. This is illustrated in Figure~\ref{tworat} where we plot the
10.52/14.74~$\mu$m ratio against the 10.52/15.46~$\mu$m ratio for various temperatures and
densities. We discussed in Section~\ref{Avalues} the role of the $a^5$F$_5$--$a^3$F$_4$ transition
in determining the relative populations of the $a^5$F and $a^3$F levels and the rather large
differences in the calculated values for the radiative decay probability for this transition. The
value calculated by \citet{Quinet1998} is about a factor of three smaller than ours and a lower
transition probability would mean that the critical density for collisional de-excitation of the
$a^5$F levels would be three times smaller. Above the critical density the relative level
populations tend to their Boltzmann values and depend only on temperature as shown by the lines in
Figure~\ref{levelsandirlines} corresponding to electron number densities of $10^7$ and
$10^8$~cm$^{-3}$. If the transition probability of \citet{Quinet1998} were adopted, these high
density limit values of the ratios would be reached at a density three times lower.

\section{Conclusions} \label{Conclusions}

In this paper, the Co\II\ forbidden transitions between the fifteen lowest energy levels of
singly-ionised cobalt, Co$^+$, have been investigated. Radiative transition probabilities and, for
the first time for this ion, collision strengths for excitation by electrons and their
thermally-averaged values based on a Maxwell-Boltzmann electron energy distribution have been
calculated. An elaborate atomic target was used to perform the scattering calculations using the
R-matrix method in the Breit-Pauli approximation and intermediate coupling scheme.

We have compared the radiative transition probabilities with those from three previous calculations
by \citet{NussbaumerS1988}, \citet{Quinet1998} and \citet{RaassenPU1998}, and briefly discussed the
consequences of the differences seen for the spectroscopy of the infrared lines. Although there is
good agreement between our data and the previous works, there are also some significant
differences, particularly for the important $a^5$F$_5$--$a^3$F$_4$ transition. This probability is
clearly very sensitive to the choice of configurations and orbitals, and we have argued that our
basis and orbitals give the most reliable value for this probability.

The diagnostic potential of the infrared lines is illustrated by showing that the three lines at
10.52, 14.74 and 15.46~$\mu$m can potentially be used to measure the electron density and
temperature of the emitting region. These lines are of particular use in the study of supernovae.

\section{Acknowledgments}

PJS and CJZ acknowledge financial support from the Atomic Physics for Astrophysics Project (APAP)
funded by the Science and Technology Facilities Council (STFC). The present collaboration benefited
from visits to Meudon by PJS from 1998 to 2013, with support provided by the Observatoire de Paris
and by the Universit\'{e} Paris 7. The hospitality of the Observatoire de Paris was much
appreciated.

\onecolumn

\begin{table}
\caption{Transition probabilities in s$^{-1}$ among the energetically lowest 15 levels (see
Table~\ref{levellist} for level indexing) as obtained from the current work (CW),
\citet{NussbaumerS1988} (NS), \citet{Quinet1998} (Q) and \citet{RaassenPU1998} (RPU) where $i$ and
$j$ refer to the lower and upper levels respectively. Only transition probabilities that are at
least 1\% of the total probability from a given upper level are listed. The powers of 10 by which
the number is to be multiplied are given in brackets.} \centering
\begin{tabular*}{\textwidth}{@{\extracolsep{\fill}}lrccccrrcccc}
\noalign{\hrule}
  \multicolumn{2}{c}{Transition}  &  CW & NS &  Q & RPU & \multicolumn{2}{c}{\hspace{1.5cm} Transition} & CW & NS  &  Q & RPU \\
$j$  & $i$  &  &   &  &  & & $j$  & $i$  &   &  &   \\
\noalign{\hrule}
   2 &   1 &   2.23(-2) & 2.23(-2) & 2.21(-2) & 2.24(-2) & \hspace{1.0cm} 10 &   9 &   1.88(-2) & 1.87(-2) & 1.67(-2)  & 1.88(-2) \\
   3 &   2 &   9.71(-3) & 9.73(-3) & 9.88(-3) & 9.73(-3) & 11 &   2 &   2.73(-2) & 1.87(-2) & 3.16(-2)  & 1.72(-2) \\
   4 &   1 &   9.13(-6) & 4.90(-6) & 3.09(-6) & --- &  11 &   3 &   2.81(-2) & 3.09(-2) & 4.06(-2)  & 2.57(-2) \\
   5 &   4 &   1.24(-2) & 1.23(-2) & 1.03(-2) & 1.24(-2) & 11 &   7 &   4.61(-3) & 5.30(-3) & 4.34(-3)  & 4.39(-3) \\
   6 &   5 &   1.08(-2) & 1.08(-2) & 9.27(-3) & 1.09(-2) & 11 &   8 &   1.41(-2) & 1.69(-2) & 1.34(-2)  & 1.35(-2) \\
   7 &   6 &   5.72(-3) & 5.72(-3) & 5.00(-3) & 5.74(-3) & 11 &  10 &   7.82(-3) & 8.32(-3) & 7.26(-3)  & 8.19(-3) \\
   8 &   3 &   4.33(-5) & ---      & ---      & ---      &  12 &   1 &   7.61(-3) & --- & 9.74(-3)  & --- \\
   8 &   7 &   1.78(-3) & 1.78(-3) & 1.57(-3) & 1.79(-3) & 12 &   2 &   1.67(-1) & --- & 1.74(-1)  & --- \\
   9 &   1 &   3.36(-2) & 3.85(-2) & 5.36(-2) & 3.43(-2) & 12 &   3 &   7.99(-2) & --- & 8.16(-2)  & 1.27(-3) \\
   9 &   2 &   3.98(-3) & 4.91(-3) & 6.33(-3) & 4.01(-3) & 13 &   1 &   3.34(-2) & --- & 4.98(-2)  & --- \\
   9 &   4 &   2.90(-2) & 2.89(-2) & 2.66(-2) & 2.81(-2) & 13 &   2 &   6.98(-2) & --- & 6.70(-2)  & --- \\
   9 &   5 &   3.27(-3) & 2.82(-3) & 2.98(-3) & 3.07(-3) & 13 &   3 &   1.82(-2) & --- & 1.63(-2)  & --- \\
   9 &   6 &   5.21(-3) & 4.28(-3) & 4.81(-3) & 5.05(-3) & 13 &   12 &   3.01(-2) & --- & 2.96(-2)  & --- \\
  10 &   1 &   1.45(-2) & 1.63(-3) & 2.30(-2) & 1.51(-2) & 14 &   2 &   2.92(-2) & --- & 4.20(-2)  & --- \\
  10 &   2 &   2.22(-2) & 2.58(-2) & 3.47(-2) & 2.25(-2) & 14 &   3 &   1.19(-2) & --- & 1.68(-2)  & --- \\
  10 &   3 &   5.60(-3) & 6.97(-3) & 8.74(-3) & 5.65(-3) & 14 &   11 &   1.99(-3) & --- & ---   & --- \\
  10 &   5 &   4.43(-3) & 6.41(-3) & 4.17(-3) & 3.97(-3) & 14 &   12 &   2.78 (-2) & --- & 2.55(-2)  & --- \\
  10 &   6 &   4.95(-3) & 4.91(-3) & 4.53(-3) & 4.57(-3) & 15 &   3 &   3.70(-2) & --- & 5.34(-2)  & --- \\
  10 &   7 &   1.04(-2) & 1.04(-2) & 9.59(-3) & 9.65(-3) &   \\
\noalign{\hrule}
\end{tabular*}
\label{Avalues2}
\end{table}


\clearpage

{\small
\begin{longtable}{cccccccccccccccc}
\caption{Thermally-averaged collision strengths among the 15 energetically lowest levels as a
function of log$_{10}$ of temperature in K where $i$ and $j$ refer to the index of the lower and
upper level respectively (see
Table~\ref{levellist} for indexing). \label{upstable}} \vspace{0cm}\\
\hline
        & $i$  &      $j$      &                                                                                                                                        \multicolumn{ 13}{c}{log$_{10}T$} \\
\cline{4-16}
       &   &           &          2.0 &        2.2 &        2.4 &        2.6 &        2.8 &          3.0 &        3.2 &        3.4 &        3.6 &        3.8 &          4.0 &        4.2 &        4.4 \\
\hline
\endfirsthead
\caption[]{continued.}\\
\hline
         & $i$ &       $j$     &                                                                                                                                        \multicolumn{ 13}{c}{log$_{10}T$} \\
\cline{4-16}
        &  &           &          2.0 &        2.2 &        2.4 &        2.6 &        2.8 &          3.0 &        3.2 &        3.4 &        3.6 &        3.8 &          4.0 &        4.2 &        4.4 \\
\hline
\endhead
\hline
\endfoot
 &  1 &  2 & 4.250 & 4.079 & 3.969 & 3.899 & 3.760 & 3.519 & 3.308 & 3.297 & 3.412 & 3.420 & 3.235 & 2.925 & 2.563 \\
 &  1 &  3 & 1.449 & 1.320 & 1.189 & 1.064 & 0.947 & 0.845 & 0.781 & 0.775 & 0.802 & 0.807 & 0.772 & 0.709 & 0.636 \\
 &  1 &  4 & 3.202 & 3.326 & 3.404 & 3.400 & 3.369 & 3.405 & 3.512 & 3.579 & 3.495 & 3.242 & 2.891 & 2.537 & 2.209 \\
 &  1 &  5 & 2.292 & 2.136 & 1.988 & 1.918 & 1.978 & 2.138 & 2.285 & 2.314 & 2.191 & 1.955 & 1.684 & 1.447 & 1.251 \\
 &  1 &  6 & 0.674 & 0.751 & 0.834 & 0.937 & 1.054 & 1.149 & 1.186 & 1.150 & 1.049 & 0.909 & 0.767 & 0.656 & 0.572 \\
 &  1 &  7 & 0.235 & 0.241 & 0.271 & 0.319 & 0.366 & 0.395 & 0.401 & 0.385 & 0.349 & 0.301 & 0.254 & 0.220 & 0.196 \\
 &  1 &  8 & 0.037 & 0.039 & 0.046 & 0.057 & 0.071 & 0.081 & 0.084 & 0.081 & 0.073 & 0.063 & 0.053 & 0.047 & 0.043 \\
 &  1 &  9 & 2.091 & 2.135 & 2.169 & 2.196 & 2.212 & 2.213 & 2.192 & 2.142 & 2.065 & 1.981 & 1.941 & 1.975 & 2.004 \\
 &  1 & 10 & 0.596 & 0.589 & 0.586 & 0.591 & 0.607 & 0.625 & 0.633 & 0.626 & 0.609 & 0.591 & 0.594 & 0.630 & 0.662 \\
 &  1 & 11 & 0.143 & 0.146 & 0.150 & 0.156 & 0.161 & 0.161 & 0.155 & 0.146 & 0.136 & 0.127 & 0.128 & 0.139 & 0.149 \\
 &  1 & 12 & 0.679 & 0.726 & 0.809 & 0.926 & 1.033 & 1.084 & 1.080 & 1.055 & 1.029 & 1.003 & 0.973 & 0.942 & 0.915 \\
 &  1 & 13 & 0.641 & 0.654 & 0.654 & 0.637 & 0.618 & 0.620 & 0.642 & 0.672 & 0.706 & 0.741 & 0.769 & 0.783 & 0.773 \\
 &  1 & 14 & 0.253 & 0.251 & 0.241 & 0.228 & 0.218 & 0.216 & 0.220 & 0.228 & 0.243 & 0.268 & 0.297 & 0.320 & 0.328 \\
 &  1 & 15 & 0.074 & 0.070 & 0.065 & 0.062 & 0.061 & 0.063 & 0.066 & 0.069 & 0.072 & 0.077 & 0.086 & 0.093 & 0.095 \\
 &  2 &  3 & 4.806 & 4.372 & 3.886 & 3.409 & 2.967 & 2.590 & 2.361 & 2.357 & 2.475 & 2.513 & 2.399 & 2.182 & 1.919 \\
 &  2 &  4 & 1.081 & 1.087 & 1.068 & 1.026 & 0.977 & 0.946 & 0.936 & 0.924 & 0.880 & 0.799 & 0.700 & 0.608 & 0.528 \\
 &  2 &  5 & 2.013 & 2.021 & 1.980 & 1.902 & 1.828 & 1.790 & 1.783 & 1.762 & 1.680 & 1.527 & 1.337 & 1.157 & 0.997 \\
 &  2 &  6 & 1.445 & 1.437 & 1.434 & 1.435 & 1.456 & 1.500 & 1.543 & 1.541 & 1.463 & 1.318 & 1.147 & 0.993 & 0.863 \\
 &  2 &  7 & 0.912 & 0.868 & 0.837 & 0.839 & 0.875 & 0.926 & 0.963 & 0.960 & 0.906 & 0.811 & 0.704 & 0.612 & 0.538 \\
 &  2 &  8 & 0.258 & 0.270 & 0.296 & 0.340 & 0.390 & 0.431 & 0.453 & 0.452 & 0.426 & 0.381 & 0.331 & 0.290 & 0.257 \\
 &  2 &  9 & 0.733 & 0.795 & 0.835 & 0.847 & 0.836 & 0.810 & 0.778 & 0.743 & 0.708 & 0.674 & 0.654 & 0.654 & 0.649 \\
 &  2 & 10 & 1.314 & 1.334 & 1.317 & 1.273 & 1.225 & 1.182 & 1.137 & 1.086 & 1.031 & 0.979 & 0.957 & 0.980 & 1.006 \\
 &  2 & 11 & 0.486 & 0.484 & 0.490 & 0.509 & 0.531 & 0.545 & 0.544 & 0.531 & 0.512 & 0.495 & 0.495 & 0.518 & 0.540 \\
 &  2 & 12 & 0.448 & 0.455 & 0.497 & 0.579 & 0.672 & 0.732 & 0.749 & 0.740 & 0.723 & 0.700 & 0.676 & 0.649 & 0.617 \\
 &  2 & 13 & 0.550 & 0.554 & 0.551 & 0.544 & 0.538 & 0.544 & 0.562 & 0.581 & 0.596 & 0.607 & 0.616 & 0.619 & 0.610 \\
 &  2 & 14 & 0.370 & 0.397 & 0.409 & 0.406 & 0.397 & 0.391 & 0.389 & 0.388 & 0.392 & 0.402 & 0.418 & 0.433 & 0.436 \\
 &  2 & 15 & 0.070 & 0.064 & 0.058 & 0.054 & 0.053 & 0.055 & 0.059 & 0.065 & 0.073 & 0.084 & 0.094 & 0.100 & 0.103 \\
 &  3 &  4 & 0.173 & 0.166 & 0.158 & 0.149 & 0.141 & 0.137 & 0.139 & 0.139 & 0.132 & 0.118 & 0.101 & 0.086 & 0.075 \\
 &  3 &  5 & 0.783 & 0.771 & 0.747 & 0.712 & 0.673 & 0.642 & 0.623 & 0.606 & 0.574 & 0.520 & 0.455 & 0.394 & 0.341 \\
 &  3 &  6 & 1.197 & 1.183 & 1.151 & 1.100 & 1.051 & 1.028 & 1.030 & 1.030 & 0.992 & 0.908 & 0.801 & 0.698 & 0.606 \\
 &  3 &  7 & 1.208 & 1.174 & 1.121 & 1.072 & 1.054 & 1.077 & 1.117 & 1.131 & 1.090 & 0.995 & 0.876 & 0.765 & 0.669 \\
 &  3 &  8 & 0.575 & 0.577 & 0.592 & 0.629 & 0.693 & 0.769 & 0.829 & 0.848 & 0.815 & 0.742 & 0.652 & 0.571 & 0.501 \\
 &  3 &  9 & 0.334 & 0.342 & 0.338 & 0.321 & 0.294 & 0.263 & 0.233 & 0.207 & 0.185 & 0.167 & 0.156 & 0.152 & 0.150 \\
 &  3 & 10 & 0.748 & 0.770 & 0.768 & 0.743 & 0.707 & 0.672 & 0.638 & 0.605 & 0.572 & 0.544 & 0.530 & 0.533 & 0.533 \\
 &  3 & 11 & 0.839 & 0.827 & 0.824 & 0.840 & 0.869 & 0.891 & 0.892 & 0.876 & 0.850 & 0.820 & 0.811 & 0.838 & 0.864 \\
 &  3 & 12 & 0.265 & 0.283 & 0.318 & 0.381 & 0.453 & 0.501 & 0.515 & 0.510 & 0.497 & 0.483 & 0.475 & 0.465 & 0.441 \\
 &  3 & 13 & 0.465 & 0.467 & 0.453 & 0.433 & 0.417 & 0.413 & 0.422 & 0.434 & 0.442 & 0.448 & 0.454 & 0.457 & 0.452 \\
 &  3 & 14 & 0.254 & 0.262 & 0.272 & 0.279 & 0.283 & 0.284 & 0.285 & 0.284 & 0.288 & 0.298 & 0.311 & 0.320 & 0.321 \\
 &  3 & 15 & 0.178 & 0.168 & 0.158 & 0.149 & 0.143 & 0.141 & 0.141 & 0.141 & 0.143 & 0.147 & 0.152 & 0.158 & 0.160 \\
 &  4 &  5 & 9.862 & 9.916 & 9.923 & 9.890 & 9.662 & 9.214 & 8.727 & 8.312 & 7.876 & 7.307 & 6.712 & 6.308 & 5.996 \\
 &  4 &  6 & 0.395 & 0.419 & 0.485 & 0.581 & 0.666 & 0.703 & 0.686 & 0.630 & 0.549 & 0.469 & 0.440 & 0.499 & 0.594 \\
 &  4 &  7 & 0.127 & 0.156 & 0.196 & 0.239 & 0.274 & 0.285 & 0.271 & 0.240 & 0.202 & 0.167 & 0.150 & 0.162 & 0.187 \\
 &  4 &  8 & 0.073 & 0.075 & 0.082 & 0.095 & 0.105 & 0.104 & 0.093 & 0.077 & 0.061 & 0.049 & 0.042 & 0.045 & 0.051 \\
 &  4 &  9 &18.971 &18.883 &18.780 &18.640 &18.447 &18.175 &17.721 &16.898 &15.540 &13.714 &11.799 &10.115 & 8.614 \\
 &  4 & 10 & 0.276 & 0.240 & 0.211 & 0.190 & 0.174 & 0.161 & 0.150 & 0.145 & 0.145 & 0.157 & 0.205 & 0.294 & 0.375 \\
 &  4 & 11 & 0.044 & 0.052 & 0.056 & 0.053 & 0.047 & 0.041 & 0.036 & 0.034 & 0.034 & 0.036 & 0.046 & 0.066 & 0.085 \\
 &  4 & 12 & 0.298 & 0.297 & 0.301 & 0.304 & 0.303 & 0.297 & 0.286 & 0.272 & 0.258 & 0.244 & 0.230 & 0.218 & 0.201 \\
 &  4 & 13 & 0.556 & 0.548 & 0.549 & 0.554 & 0.558 & 0.558 & 0.553 & 0.545 & 0.534 & 0.521 & 0.508 & 0.494 & 0.468 \\
 &  4 & 14 & 0.017 & 0.021 & 0.025 & 0.027 & 0.027 & 0.027 & 0.025 & 0.024 & 0.023 & 0.023 & 0.027 & 0.033 & 0.037 \\
 &  4 & 15 & 0.036 & 0.029 & 0.023 & 0.018 & 0.014 & 0.011 & 0.010 & 0.009 & 0.008 & 0.007 & 0.008 & 0.009 & 0.009 \\
 &  5 &  6 & 7.160 & 7.608 & 8.160 & 8.611 & 8.810 & 8.785 & 8.694 & 8.618 & 8.446 & 8.035 & 7.446 & 6.867 & 6.285 \\
 &  5 &  7 & 0.499 & 0.491 & 0.506 & 0.548 & 0.587 & 0.596 & 0.570 & 0.521 & 0.456 & 0.391 & 0.368 & 0.419 & 0.499 \\
 &  5 &  8 & 0.138 & 0.168 & 0.206 & 0.240 & 0.256 & 0.251 & 0.232 & 0.205 & 0.175 & 0.148 & 0.136 & 0.151 & 0.178 \\
 &  5 &  9 & 8.909 & 8.860 & 8.811 & 8.728 & 8.556 & 8.267 & 7.865 & 7.337 & 6.647 & 5.831 & 5.067 & 4.493 & 4.005 \\
 &  5 & 10 & 8.662 & 8.640 & 8.655 & 8.684 & 8.690 & 8.634 & 8.451 & 8.055 & 7.388 & 6.496 & 5.563 & 4.743 & 4.022 \\
 &  5 & 11 & 0.108 & 0.110 & 0.110 & 0.106 & 0.099 & 0.090 & 0.083 & 0.080 & 0.080 & 0.087 & 0.117 & 0.173 & 0.226 \\
 &  5 & 12 & 0.262 & 0.251 & 0.243 & 0.234 & 0.226 & 0.216 & 0.207 & 0.197 & 0.187 & 0.177 & 0.168 & 0.161 & 0.150 \\
 &  5 & 13 & 0.357 & 0.346 & 0.338 & 0.332 & 0.327 & 0.323 & 0.318 & 0.312 & 0.303 & 0.296 & 0.296 & 0.301 & 0.296 \\
 &  5 & 14 & 0.275 & 0.276 & 0.277 & 0.275 & 0.273 & 0.270 & 0.267 & 0.262 & 0.255 & 0.248 & 0.242 & 0.239 & 0.229 \\
 &  5 & 15 & 0.020 & 0.016 & 0.013 & 0.010 & 0.009 & 0.009 & 0.009 & 0.009 & 0.008 & 0.009 & 0.010 & 0.012 & 0.013 \\
 &  6 &  7 & 5.975 & 6.146 & 6.457 & 6.876 & 7.209 & 7.337 & 7.361 & 7.372 & 7.285 & 6.978 & 6.489 & 5.970 & 5.433 \\
 &  6 &  8 & 0.360 & 0.382 & 0.423 & 0.466 & 0.489 & 0.481 & 0.449 & 0.403 & 0.350 & 0.300 & 0.283 & 0.323 & 0.386 \\
 &  6 &  9 & 2.461 & 2.482 & 2.488 & 2.472 & 2.411 & 2.293 & 2.131 & 1.939 & 1.723 & 1.503 & 1.342 & 1.280 & 1.243 \\
 &  6 & 10 & 9.081 & 9.062 & 9.032 & 8.955 & 8.811 & 8.598 & 8.290 & 7.826 & 7.143 & 6.270 & 5.377 & 4.611 & 3.940 \\
 &  6 & 11 & 3.168 & 3.171 & 3.167 & 3.161 & 3.148 & 3.111 & 3.027 & 2.872 & 2.629 & 2.318 & 2.011 & 1.765 & 1.552 \\
 &  6 & 12 & 0.282 & 0.254 & 0.232 & 0.211 & 0.191 & 0.174 & 0.161 & 0.150 & 0.140 & 0.129 & 0.121 & 0.113 & 0.104 \\
 &  6 & 13 & 0.232 & 0.235 & 0.232 & 0.226 & 0.218 & 0.211 & 0.203 & 0.193 & 0.182 & 0.172 & 0.171 & 0.174 & 0.172 \\
 &  6 & 14 & 0.309 & 0.323 & 0.331 & 0.329 & 0.320 & 0.310 & 0.301 & 0.291 & 0.279 & 0.268 & 0.260 & 0.256 & 0.247 \\
 &  6 & 15 & 0.076 & 0.078 & 0.076 & 0.073 & 0.071 & 0.070 & 0.069 & 0.068 & 0.066 & 0.064 & 0.064 & 0.064 & 0.063 \\
 &  7 &  8 & 3.196 & 3.382 & 3.658 & 4.001 & 4.320 & 4.516 & 4.620 & 4.686 & 4.669 & 4.496 & 4.200 & 3.892 & 3.571 \\
 &  7 &  9 & 0.183 & 0.212 & 0.230 & 0.238 & 0.236 & 0.220 & 0.193 & 0.160 & 0.131 & 0.112 & 0.115 & 0.148 & 0.186 \\
 &  7 & 10 & 4.616 & 4.550 & 4.480 & 4.380 & 4.237 & 4.059 & 3.851 & 3.594 & 3.261 & 2.861 & 2.476 & 2.169 & 1.902 \\
 &  7 & 11 & 5.926 & 5.922 & 5.910 & 5.894 & 5.868 & 5.805 & 5.663 & 5.388 & 4.941 & 4.351 & 3.739 & 3.209 & 2.741 \\
 &  7 & 12 & 0.192 & 0.179 & 0.166 & 0.151 & 0.135 & 0.120 & 0.108 & 0.098 & 0.090 & 0.081 & 0.074 & 0.068 & 0.061 \\
 &  7 & 13 & 0.121 & 0.127 & 0.130 & 0.131 & 0.130 & 0.127 & 0.122 & 0.114 & 0.105 & 0.097 & 0.093 & 0.093 & 0.091 \\
 &  7 & 14 & 0.316 & 0.305 & 0.294 & 0.279 & 0.263 & 0.248 & 0.237 & 0.225 & 0.213 & 0.202 & 0.196 & 0.194 & 0.188 \\
 &  7 & 15 & 0.175 & 0.163 & 0.149 & 0.136 & 0.126 & 0.121 & 0.116 & 0.112 & 0.108 & 0.103 & 0.100 & 0.098 & 0.094 \\
 &  8 &  9 & 0.037 & 0.048 & 0.057 & 0.064 & 0.069 & 0.069 & 0.064 & 0.056 & 0.047 & 0.041 & 0.040 & 0.049 & 0.059 \\
 &  8 & 10 & 0.310 & 0.305 & 0.295 & 0.273 & 0.241 & 0.201 & 0.163 & 0.130 & 0.106 & 0.094 & 0.104 & 0.137 & 0.171 \\
 &  8 & 11 & 5.681 & 5.677 & 5.681 & 5.683 & 5.674 & 5.630 & 5.508 & 5.256 & 4.833 & 4.263 & 3.660 & 3.124 & 2.645 \\
 &  8 & 12 & 0.074 & 0.077 & 0.078 & 0.074 & 0.068 & 0.061 & 0.055 & 0.050 & 0.045 & 0.041 & 0.037 & 0.034 & 0.030 \\
 &  8 & 13 & 0.057 & 0.061 & 0.064 & 0.065 & 0.065 & 0.064 & 0.061 & 0.058 & 0.053 & 0.048 & 0.046 & 0.045 & 0.043 \\
 &  8 & 14 & 0.176 & 0.174 & 0.172 & 0.167 & 0.160 & 0.151 & 0.143 & 0.134 & 0.125 & 0.117 & 0.113 & 0.111 & 0.107 \\
 &  8 & 15 & 0.114 & 0.111 & 0.107 & 0.103 & 0.100 & 0.099 & 0.098 & 0.096 & 0.093 & 0.089 & 0.086 & 0.084 & 0.080 \\
 &  9 & 10 & 3.230 & 3.180 & 3.138 & 3.100 & 3.060 & 2.997 & 2.881 & 2.696 & 2.452 & 2.204 & 2.085 & 2.170 & 2.298 \\
 &  9 & 11 & 0.138 & 0.142 & 0.149 & 0.164 & 0.185 & 0.204 & 0.215 & 0.216 & 0.216 & 0.224 & 0.261 & 0.339 & 0.420 \\
 &  9 & 12 & 0.269 & 0.337 & 0.414 & 0.481 & 0.520 & 0.521 & 0.491 & 0.451 & 0.412 & 0.378 & 0.351 & 0.331 & 0.310 \\
 &  9 & 13 & 0.536 & 0.532 & 0.538 & 0.543 & 0.540 & 0.531 & 0.521 & 0.512 & 0.507 & 0.506 & 0.516 & 0.538 & 0.547 \\
 &  9 & 14 & 0.054 & 0.060 & 0.065 & 0.069 & 0.074 & 0.079 & 0.083 & 0.088 & 0.092 & 0.098 & 0.109 & 0.122 & 0.129 \\
 &  9 & 15 & 0.013 & 0.011 & 0.009 & 0.008 & 0.008 & 0.009 & 0.010 & 0.012 & 0.013 & 0.016 & 0.019 & 0.022 & 0.024 \\
 & 10 & 11 & 2.666 & 2.682 & 2.690 & 2.686 & 2.659 & 2.593 & 2.475 & 2.304 & 2.086 & 1.858 & 1.727 & 1.750 & 1.811 \\
 & 10 & 12 & 0.229 & 0.263 & 0.315 & 0.367 & 0.397 & 0.393 & 0.364 & 0.326 & 0.292 & 0.263 & 0.240 & 0.223 & 0.205 \\
 & 10 & 13 & 0.257 & 0.259 & 0.260 & 0.258 & 0.253 & 0.250 & 0.249 & 0.248 & 0.245 & 0.245 & 0.253 & 0.271 & 0.280 \\
 & 10 & 14 & 0.396 & 0.406 & 0.408 & 0.397 & 0.379 & 0.362 & 0.349 & 0.339 & 0.330 & 0.325 & 0.327 & 0.336 & 0.341 \\
 & 10 & 15 & 0.051 & 0.050 & 0.048 & 0.046 & 0.045 & 0.046 & 0.047 & 0.049 & 0.049 & 0.050 & 0.052 & 0.056 & 0.058 \\
 & 11 & 12 & 0.135 & 0.154 & 0.190 & 0.233 & 0.261 & 0.265 & 0.249 & 0.225 & 0.203 & 0.183 & 0.166 & 0.151 & 0.134 \\
 & 11 & 13 & 0.119 & 0.123 & 0.123 & 0.118 & 0.114 & 0.114 & 0.119 & 0.123 & 0.125 & 0.126 & 0.130 & 0.136 & 0.137 \\
 & 11 & 14 & 0.254 & 0.261 & 0.262 & 0.256 & 0.247 & 0.239 & 0.234 & 0.230 & 0.226 & 0.222 & 0.227 & 0.241 & 0.250 \\
 & 11 & 15 & 0.235 & 0.221 & 0.204 & 0.188 & 0.175 & 0.168 & 0.164 & 0.160 & 0.156 & 0.152 & 0.150 & 0.152 & 0.153 \\
 & 12 & 13 & 0.791 & 0.869 & 0.926 & 0.947 & 0.941 & 0.949 & 0.992 & 1.046 & 1.084 & 1.103 & 1.104 & 1.075 & 1.004 \\
 & 12 & 14 & 0.295 & 0.335 & 0.368 & 0.383 & 0.384 & 0.384 & 0.389 & 0.399 & 0.412 & 0.423 & 0.423 & 0.412 & 0.390 \\
 & 12 & 15 & 0.140 & 0.136 & 0.129 & 0.120 & 0.115 & 0.113 & 0.114 & 0.118 & 0.123 & 0.127 & 0.126 & 0.121 & 0.112 \\
 & 13 & 14 & 0.735 & 0.729 & 0.714 & 0.679 & 0.634 & 0.598 & 0.572 & 0.553 & 0.540 & 0.531 & 0.526 & 0.521 & 0.515 \\
 & 13 & 15 & 0.313 & 0.285 & 0.251 & 0.218 & 0.193 & 0.177 & 0.168 & 0.162 & 0.156 & 0.151 & 0.146 & 0.142 & 0.141 \\
 & 14 & 15 & 0.576 & 0.490 & 0.401 & 0.324 & 0.273 & 0.244 & 0.227 & 0.216 & 0.209 & 0.208 & 0.211 & 0.213 & 0.211 \\
\hline
\end{longtable}
}

\end{document}